\documentstyle[psfig,times]{mn}
\def\H0{{\it H}$_0$}
\def\Ms{{\it M}$_\odot$}
\def\Ls{{\it L}$_\odot$}
\def\q0{{\it q}$_0$}
\def\kmps{km~s$^{-1}$}

\def\ergps{erg~s$^{-1}$}
\def\kmpspMpc{km~s$^{-1}$~Mpc$^{-1}$}
\def\Ms{{\it M}$_\odot$}

\def\nH{$N_{\rm H}$\thinspace} 

\def\psqcm{cm$^{-2}$}
\def\ergpspsqcm{erg~cm$^{-2}$~s$^{-1}$}

\def\Zs{$Z_{\odot}$}

\def\pcubcm{cm$^{-3}$}

\title[The Fe distribution in 4C+55.16] 
{The intracluster iron distribution around 4C+55.16} 
\author[K. Iwasawa et al] 
{\parbox[]{6.5in} {K. Iwasawa, A.C. Fabian, S.W. Allen and S. Ettori}\\
\\
Institute of Astronomy, Madingley Road, Cambridge CB3 0HA\\ 
}
\date{}

\begin{document}

\maketitle

\begin{abstract}
We report on the metal distribution in the intracluster medium around
the radio galaxy 4C+55.16 ($z=0.24$) observed with the Chandra X-ray
Observatory. The radial metallicity profile shows a dramatic change at
10 arcsec ($\sim 50$ kpc) in radius from half solar to twice solar at
inner radii. Also found was a plume-like feature located at $\sim 3$
arcsec to the south-west of the centre of the galaxy, which is mostly
accounted for by a strong enhancement of iron L emission. The X-ray
spectrum of the plume is characterized by the metal abundance pattern
of Type Ia supernovae (SNeIa), i.e., large ratios of Fe to
$\alpha$-elements, with the iron metallicity being unusually high at
$7.9^{+24}_{-5.0}$ solar (90 per cent error). How the plume has
been formed is not entirely clear. The inhomogeneous iron
distribution suggested in this cluster provides important clues to
understanding the metal enrichment process of the cluster medium.
\end{abstract}


\section{introduction}

The metal abundance ratio pattern is useful for understanding the
origin of heavy elements in the intracluster medium (ICM) because of
significantly different metal yields between Type Ia and II (plus Ib and Ic)
supernovae (SNeIa and SNeII) (Renzini et al 1993). Overall, SNeII appear to
play a major role in enriching the ICM, as inferred from abundance
measurements based on ASCA data (Mushotzky et al 1996; Fukazawa et al
1998). On the other hand, it has been suggested that an
iron metallicity gradient is present in the ICM, especially of
clusters with a cooling flow.
%(e.g., the Virgo cluster, Koyama, Takano
%\& Tawara 1991; Matsumoto et al 1994; B\"ohringer et al
%2001). 
Evidence for an increasing contribution from SNeIa at inner radii,
inferred from an enahancement of iron abundance relative to $\alpha$
elements, has been reported for some clusters (Finoguenov, David \&
Ponman 2000; Fukazawa et al 2000; Kaastra et al 2001). However, the iron abundance generally remains significantly
sub-solar and
rarely exceeds 1 solar even at the metallicity peaks in nearby
clusters (e.g., De Grandi \& Molendi 2001). 

Whereas the limited spatial resolution of the previous X-ray
telescopes means that studies of a metallicity gradient in ICM are
possible only for nearby clusters on a large scale, with the arcsecond
resolution of Chandra X-ray Obseratory (Weisskopf et al 1996), we are
now able to explore a cluster core at higher redshift in much more
detail.
In this Letter, we present spatially resolved high metallicity
concentration in the central part of the cluster around 4C+55.16.
A comprehensive study of the cluster emission 
will be presented elsewhere.
 
4C+55.16 is a compact powerful radio source residing in a large 
galaxy at a redshift of 0.240 (Pearson \& Readhead 1981, 1988; 
Whyborn et al 1985; Hutchings, Johnson \& Pyke 1988).
%Polatidis et al 1995;
%Lawrence et al 1996).
Recently, luminous cluster emission ($\sim 10^{45}$\ergps) 
around the radio galaxy has been
recognized through ASCA and ROSAT HRI observations (Iwasawa et al 1999).
We assume \H0 = 50 \kmpspMpc\ and \q0\ = 0.5 throughout this
paper. The angular scale is then 4.8 kpc arcsec$^{-1}$ for a redshift
of 0.24.

\section{observation and data reduction}

4C+55.16 was observed with the Chandra X-ray Observatory on 2000
October 8. The galaxy was positioned on the ACIS-S3 detector, which
was operating in Faint mode at a focal plane temperature of
$-120^{\circ}$C. The background of the CCD chip did not show
significant flare events but remained within 20 per cent fluctuation
around the mean count rate most of the total 9.07 ks exposure. The
data were reduced using software in CIAO 2.1.3 and the latest
calibration (caldb 2.7) released in August 2001.  Despite improved
quality of spectral fits relative to the previous calibration, the
redshift inferred from the Fe K line is $0.254^{+0.010}_{-0.009}$,
which we use in the spectral analysis but is higher than 0.240 from the
optical spectroscopy of the cD galaxy (Lawrence et al 1996).

Spectral analysis was performed using XSPEC version 11.  Spectral
parameters obtained from the analysis are presented with the 90 per
cent confidence region for one parameter of interest throughout this
paper, unless stated otherwise.

\section{results}

\subsection{X-ray morphology of cluster core}

% Full band Central part --- Fig. 1

\begin{figure}
\centerline{\psfig{figure=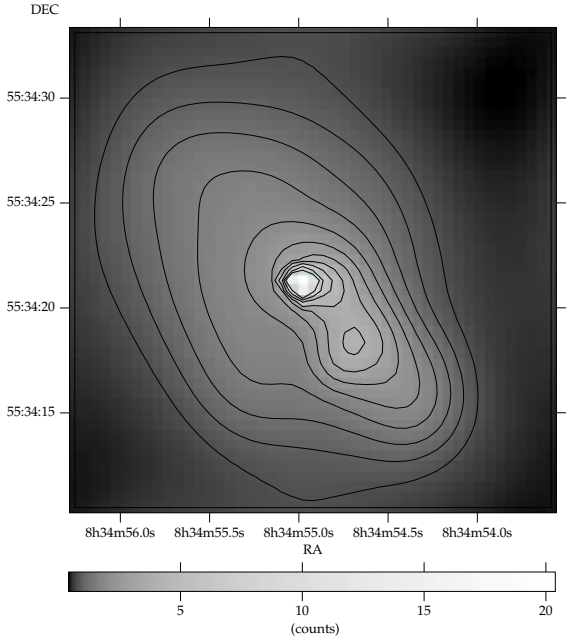,width=0.42\textwidth,angle=0}}
\caption{Central part of the X-ray emission around 4C+55.16. The image
size is $24\times 24$ arcsec. The Chandra ACIS-S 0.3--7 keV band image
has been adaptively smoothed with a $3\sigma$ gaussian
kernel. Contours are drawn at 10 equally-divided logarithmic intervals
in counts per original pixel between 4 and 40 per cent of the peak
brightness occuring at the active nucleus of the radio galaxy. A
plume-like feature is seen to the south-west.}
\end{figure}

Adaptively smoothed 0.3--7 keV image of the cluster core obtained from
the ACIS-S3 is shown in Fig. 1. An unresolved bright X-ray source is
found at the position of the nucleus of the radio galaxy.  The X-ray
cluster emission is elongated along the major axis of the optical
isophote of 4C+55.16 at a position angle of $\sim 30^{\circ}$.
A notable feature seen in the
cluster core region is a plume-like X-ray excess extending by $\sim 5$
arcsec to the SW of the nucleus. There is no obvious counterpart in
the optical and near-infrared images 
(Hutchings et al 1988; C. Crawford, priv. comm.).

\subsection{Radial variation of the ICM properties}

% Spectra of the nucleus and 8 annuli -- Fig. 2

\begin{figure}
\centerline{\psfig{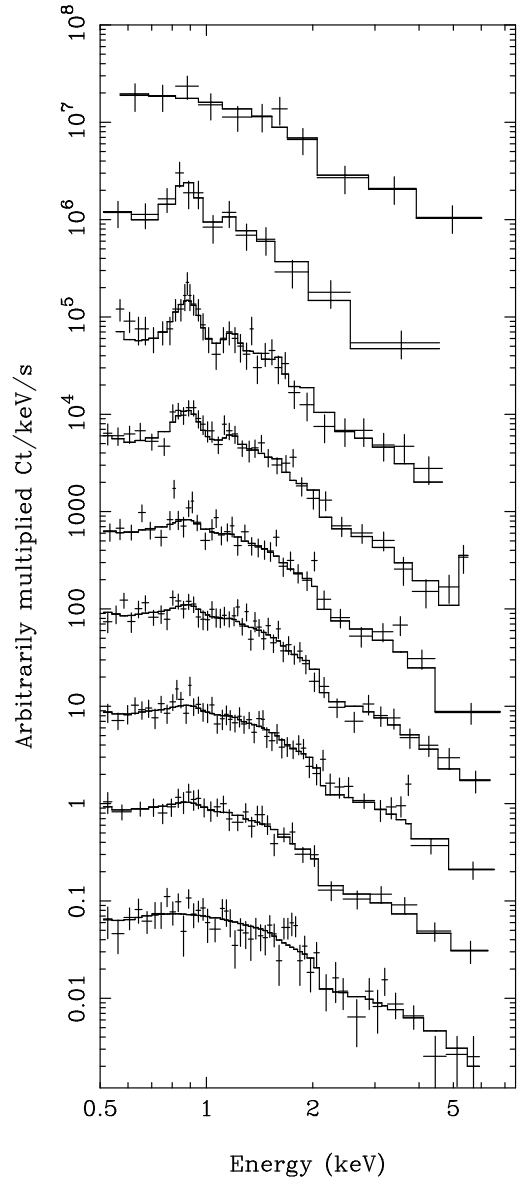}}
\caption{The ACIS-S spectra of the active nucleus and the ICM taken from
annuli in radii of 1--2.5, 2.5--5, 5--9, 9--15, 15--25, 25--40, 40--60
and 60--100 arcsec from top to bottom. Strong Fe L features are evident 
in the spectra from the inner three annuli.}
\end{figure}

% Temperature and metallicity variation -- Fig. 3

\begin{figure}
\centerline{\psfig{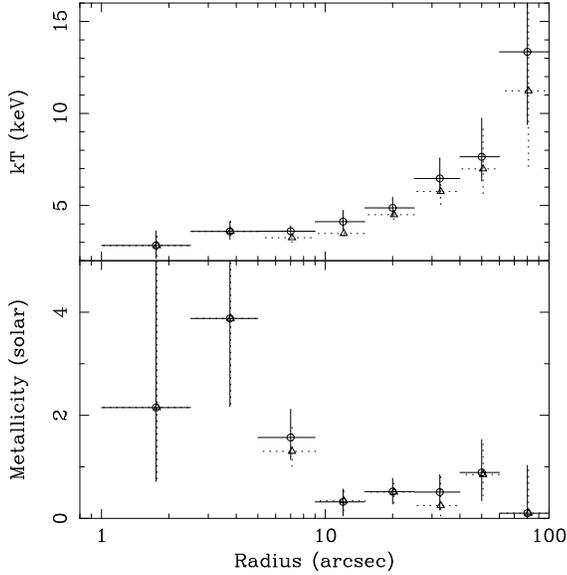}}
\caption{The radial profiles of temperature and metallicity obtained
from spectral fits to the eight annulus-spectra shown in Fig. 2 with a
single MEKAL model. Results when the spectra are modified by Galactic
absorption (circles with solid-line error bars) and variable excess
absorption is allowed (triangles with dotted-line error bars) are shown.
The error bars are of $1\sigma$. The angular scale is 4.8 kpc arcsec$^{-1}$.}
\end{figure}

% r18-2 spectrum -- Fig. 4

\begin{figure}
\centerline{\psfig{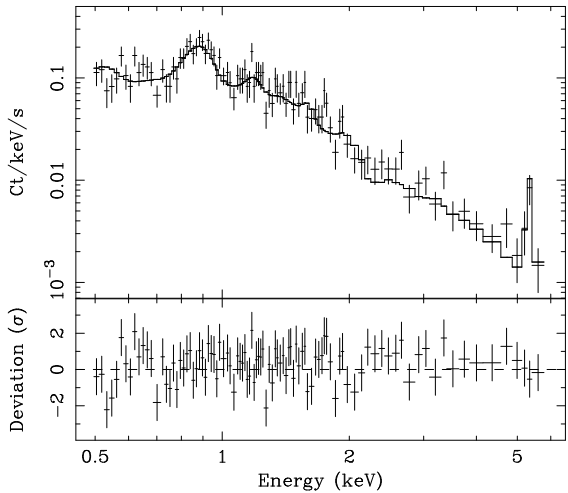}}
\caption{The ACIS-S spectrum for the central 9 arcsec region excluding the
nuclear source, fitted with a single MEKAL model modified by Galactic
absorption. A strong Fe K line is clearly detected at the observed
energy around 5.4 keV, confirming the large metallicity inferred from
the Fe L in the low energy band. The best-fit temperature and
metallicity obtained from the fit are $kT = 3.34^{+0.35}_{-0.31}$ keV
and $1.93^{+0.85}_{-0.62}$ solar, respectively.}
\end{figure}

The point source at the nucleus shows a hard X-ray spectrum, which can
be attributed naturally to nonthermal emission from the active
nucleus (Fig. 2). A power-law fit to the spectrum yields a photon index of
$1.3\pm 0.3$ with no significant excess absorption above the Galactic
value (\nH = $4.2 \times 10^{20}$\psqcm, Dickey \& Lockman 1990).  The
observed 0.5--2 keV and 2--7 keV fluxes of the nuclear source are $2.4\times
10^{-14}$\ergpspsqcm\ and $5.7\times 10^{-14}$\ergpspsqcm,
respectively. The rest-frame 2--10 keV luminosity is estimated 
to be $2\times 10^{43}$\ergps. 
%(Note that the flat spectrum is consistent 
%with a high temperature bremssstrahlung emission expected from an ADAF, unless
%it is due to complex absorption.)

Also taken were spectra from eight annuli of radii of 1--2.5, 2.5--5,
5--9, 9--15, 15--25, 25--40, 40--60, 60--100 arcsec. The background
data were taken from a region on the same detector where the cluster
emission is negligible. Each spectrum was fitted by a single MEKAL
model. For the metallicity measurement, we use solar abundance ratios
taken from Anders \& Grevesse (1989). The ACIS spectra and fitting
results on temperature and metallicity are shown in Fig. 2 and Fig. 3.
Excess absorption is not significant in most cases. However, it
affects the temperature measurements and a few times $10^{20}$\psqcm\
could still be within the calibration uncertainty. We therefore
present results in Fig. 3 both for when Galactic absorption is assumed
and variable excess absorption is allowed.

The temperature of the cluster gas shows a monotonic decrease towards
the centre. It is noted that flattening of the temperature gradient is
not seen up to 100 arcsec (480 kpc) in radius, which is rather
unusual. A similarly high temperature is obtained for the outermost
annulus when background is extracted from the blank sky data provided
by Chandra Science Center.

The metal abundance profile shows a dramatic increase at a radius
around 10 arcsec.  Within this radius, the metallicity significantly
exceeds solar whilst it is about half solar at the outer radii.
A strong Fe L feature around 0.9 keV (in the observed frame), by which
metallicity is primarily derived, is evident in the spectra from the
inner three regions (see Fig. 2). Although individual spectra do not
have sufficient counts in the Fe K band, the integrated spectrum within
a radius of 9 arcsec (excluding the nuclear source) shows a clear
strong Fe K$\alpha$ line, supporting the high metallicity implied from
the Fe L emission (Fig. 4). 

%Since these Fe L and Fe K$\alpha$ lines
%are subject to mild resonant scattering in the dense cluster core, the
%intrinsic metallicity could be slightly higher.

\subsection{Fe-L plume}

% Fe L image --- Fig. 5 (B&W version)

\begin{figure}
\centerline{\psfig{figure=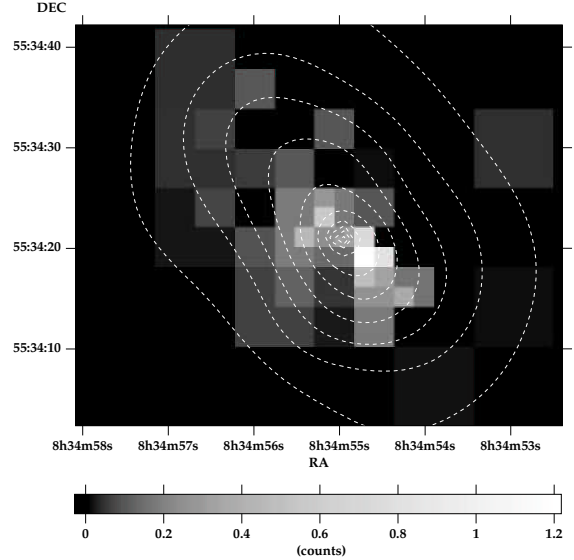,width=0.42\textwidth,angle=0}}
\caption{The continuum-subtracted Fe-L emission image of the central
part of the cluster. The 0.78--1.0 keV (in the observed frame) image
has been adaptively binned so that each pixel contains at least 9
counts, and then the continuum for each pixel estimated from the
1.3--2 keV image was subtracted (the continuum level is 22 per cent at
the Fe L peak). The colour bar shows surface brightness in counts per
original pixel. Overlaid in contours is the adaptively smoothed 1.8--5
keV band image where the continuum is a major component and is free
from Fe-L emission. The contours are drawn at 10 equally-divided
logarithmic intervals of surface brightness from 2 percent of the peak
brightness. Strong excess iron-L emission is seen at around $\sim 3$
arcsec to the SW of the galaxy nucleus and accounts most of the plume
feature, which is absent in the ICM represented by the continuum
image.}
\end{figure}

The X-ray plume located near the nucleus (see Fig. 1) is found to show
particularly strong Fe L emission.  The continuum-subtracted Fe L
image, made from the 0.78--1.0 keV (0.95--1.25 keV in the galaxy
frame) image, shows strong excess at the plume region (Fig. 5). Since
the Fe L-free band (e.g., 1.8--5 keV) image shows a smooth elliptical
isophoto centred on the galaxy nucleus with no particular excess at
the plume, the plume feature is mostly accounted for by the excess Fe
L emission. This also indicates that the ICM itself is not structured,
unlike the disturbed X-ray morphology resulted from displaced ICM by
radio jets seen in nearby clusters such as Hydra A (McNamara et al
2000) and Perseus cluster (Fabian et al 2000). In 4C+55.16, most of
the radio emission originates from the VLBI-scale compact core and the
small radio lobe, aligned approximately perpendicular to the plume,
carries only small power (Whyborn et al 1985).

% The plume spectrum -- Fig. 6

\begin{figure}
\centerline{\psfig{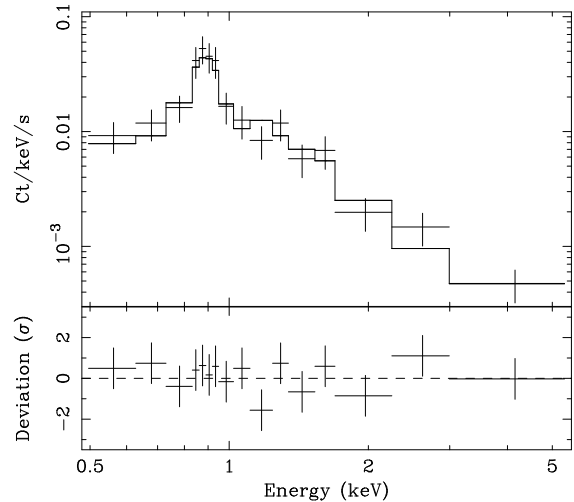}}
\caption{The ACIS-S spectrum of the plume region fitted with the VMEKAL
model with the SNIa abundance pattern (see Table 1). Note the strong
Fe L emission around 0.9 keV.}
\end{figure}

% C-shaped vs Steel Ratio and chi-sq curve --- Fig. 7

\begin{figure}
\centerline{\psfig{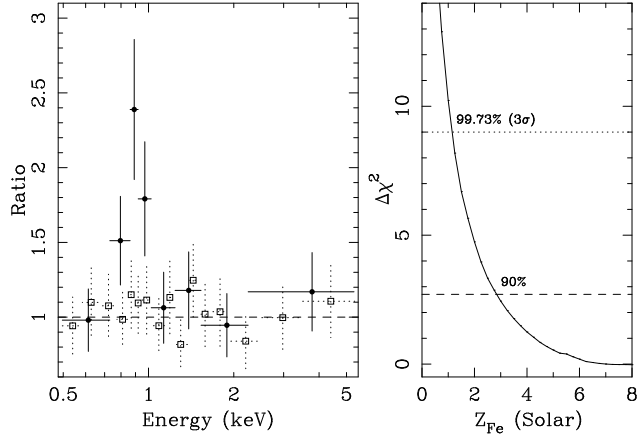}}
\caption{The left panel shows a ratio plot of the spectra from the
C-shaped region (open squares) and the plume (filled circles) relative
to the best-fit MEKAL model for the C-shaped region (Table 1),
demonstrating excess Fe L emission of the plume spectrum. The plume
data have been normalized so that the data excluding the Fe L band
match the model. The data have been rebinned further for clarity. The
right panel shows the $\chi^2$ curve for the metallicity of the plume
spectrum when fitting with the SNIa abundance pattern (Table 1). This
shows that the metallicity exceeds 1 solar at $3\sigma$ level.}
\end{figure}

The spectrum of the Fe L plume (Fig. 6) is investigated. The spectral data were
collected from a rectangular region of $3\times 4.5$arcsec with a
position angle $0^{\circ}$, centred on the plume, which covers most of
the SW quarter of the 1--5 arcsec annulus. We also take a spectrum
from the rest of the annulus (``C-shaped region'': approximately
three-quarter of the 1--5 arcsec annulus covering position angles from
$280^{\circ}$ to $180^{\circ}$ to avoid the nucleus and plume), for
comparison. 
The strong Fe L emission from the plume peaks at an energy of $0.89\pm 0.02$
keV, very similar to that ($0.89\pm 0.03$ keV) of the C-shaped region
spectrum, suggesting similar temperatures. 
Therefore the enhancement of Fe L emission seen in the
plume is more consistent with a metallicity effect than the presence
of cool gas.

Fitting the C-shaped region spectrum by a single MEKAL model gives a
good fit with a temperature of $kT = 3.9^{+1.0}_{-0.8}$ keV and
$3.4^{+11}_{-2.1}$ solar metallicity (Table 1).

The metallicity inferred for the C-shaped region spectrum is already
unusually high for ICM.  The even stronger Fe L emission seen in the
plume (Fig. 7) could be a result from selective enrichment of iron, presumably
by SNeIa. Since our short exposure spectrum of the plume (the
integrated count is $\sim 170$) does not have sufficient quality for
fitting emission lines from individual elements, we tried three
different metal abundance patterns appropriate for a) Solar
photosphere; b) SNII; and c) SNIa, in order to see which pattern
reproduces the observed spectrum best. MEKAL is used for the solar
abundance case, while the enrichment patterns by the two supernova
types are implemented with the variable abundance version of MEKAL
(VMEKAL), using the abundance ratios relative to the solar values
given in Dupke \& Arnaud (2001), based on the stellar yields for SNII
(Nomoto et al 1997a) and for the deflagration model (W7) of SNIa
(Nomoto, Thielmann \& Yokoi 1984; Nomoto et al 1997b).

The results of spectral fitting are shown in Table 1. The SNeII type
abundance pattern can be ruled out. The SNIa abundance pattern gives a
significantly better fit over that of the solar abundance
($\Delta\chi^2\sim 7$ for 13 degrees of freedom for both fits). 
The SNIa abundance model gives a temperature
$kT\simeq 3.4$ keV, similar to the surrounding C-shaped region, and
$\simeq 8$ solar iron metallicity (as shown in Fig. 7, the iron
metallicity exceeds 1 solar at $3\sigma $ confidence level).

More complex models, e.g., two-temperature model and cooling-flow
model, are not required by the data. 
The inclusion of extra components may be needed, which
could alter the metallicity measurements if the data quality is much
better, but it would usually push the metallicity value even higher.

The model of the SNIa abundance pattern is also examined for the
C-shaped region spectrum. It gives $2.3^{+1.3}_{-1.2}$ solar
metallicity, but does not improve the fit over the solar abundance
model, unlike for the plume spectrum (see Table 1).

% Table 1 -- Fitting results to Steel

\begin{table}
\begin{center}
\caption{Results of spectral fits to the ACIS-S spectra of the plume
and ``C-shaped'' regions. Three metal abundance patterns appropriate
for (a) the solar photosphere; (b) SNIa; and (c) SNII were tested for
the plume spectrum. All the fits include Galactic absorption. The
number ratios of each element to iron, relative to the solar ratio,
are as follows for the two types of supernovae, as given in Dupke \&
Arnaud (2001): (O,Mg)=0.035Fe, Ne=0.006Fe, (Si,S,Ar,Ca)=0.5Fe,
Ni=4.8Fe for a SNIa; and (O,Mg,Si)=3.7Fe, (Ne,S)=2.5Fe,
(Ar,Ca,Ni)=1.7Fe for a SNII.  $^{\ast }K$ is the normalization of the
model in unit of $10^{-19}/(4\pi D_{\rm L}^2)\int n_{\rm e}n_{\rm
H}dV$, where $D_{\rm L}$ is the luminosity distance of the source
($4.7\times 10^{27}$ cm), $n_{\rm e}$ and $n_{\rm H}$ are densities of
electron and hydrogen, respectively in cm$^{-3}$. The errors quoted
are of 90 per cent confidence region for one parameter of
interest. $^{\dagger}$This value is for the solar abundance
pattern. $Z_{\rm Fe}=2.3^{+1.3}_{-1.2}$ solar with a comparable $\chi^2$ is
obtained when the SNIa pattern is employed.}
\begin{tabular}{lcccc}
Abundance & $kT$ & $Z_{\rm Fe}$ & $K$ & $\chi^2$/dof \\
pattern & keV & \Zs & $\ast $ & \\[5pt]
%(a) Solar & $3.0^{+1.1}_{-0.8}$ & $9.1_{-6.8}^{\dagger}$ 
(a) Solar & $3.0^{+0.9}_{-0.7}$ & $>4.3$ 
& 2.1 & 14.2/13 \\[5pt]
%(b) SNIa & $3.4^{+1.5}_{-1.0}$ & $5.6^{+16.9}_{-3.4}$ & 6.0 
(b) SNIa & $3.4^{+1.0}_{-0.8}$ & $7.9^{+24}_{-5.0}$ & 5.4 
& 7.9/13 \\[5pt]
(c) SNII   & 3.1 & 0.7 & 7.3 & 32.7/13 \\[5pt]
C-shaped & $4.0^{+1.0}_{-0.8}$ & $^{\dagger}3.4^{+10}_{-2.1}$ &
14 & 12.2/29 \\
region & & & & \\
\end{tabular}
\end{center}
\end{table}

\section{discussion}

We have found a striking increase in metallicity in the cluster core
around 4C+55.16 through iron emission measurements. Metallicity
increases of a lesser degree have been found earlier towards the
centres of several nearby clusters (up to 0.5 solar, De Grandi \&
Molendi 2001). Centaurus and Virgo clusters show peak metallicity of $\sim
1.5$ solar (Ikebe et al 1999; Sanders \& Fabian 2001) and $\sim 0.8$ solar
(B\"ohringer et al 2001), respectively. S\'ersic
159-03 shows an iron-rich region (0.76 solar) displaced from
the centre by $\sim 30$ kpc (Kaastra et al 2001). Our results for the
higher redshift cluster 4C+55.16 are more dramatic than any of these,
with the cluster core being about twice solar abundance and the plume
is apparently three times richer in iron. No mechanism other than a
metallicity effect appears to be plausible: photoionization by the
active nucleus, for instance, has a problem with orientation, i.e.,
the radio axis is perpendicular to the plume; and non-equilibrium
ionization of the gas, which could be induced by a merger, would last
for less than a million years and thus is unlikely to be observed in
the absence of any optical signature of a merger.

From the temperature, metallicity, and normalization of the VMEKAL
thermal spectrum obtained from the fit to the plume spectrum with the
SNIa abundance pattern, and its assumed volume, $V = 14.4\times
14.4\times 21.6$ kpc$^3$, we obtain gas densities, $n_{\rm gas}\approx
0.177$ \pcubcm, and gas mass, $M_{\rm gas}=\mu m_{\rm p} n_{\rm gas} V
\approx 1.2\times 10^{10}$\Ms. The iron mass, $M_{\rm Fe}\approx
2.5\times 10^8$\Ms, the emissivity of the gas, $\epsilon\approx
3.3\times 10^{-25}$ erg\thinspace s$^{-1}$\thinspace cm$^{-3}$, and
its cooling time, $t_{\rm cool}=(3/2)kTn_{\rm gas}/\epsilon\approx
1.4\times 10^8$ yr are derived.

The metal-enrichment of the plume region clearly needs iron-rich
sources, such as SNeIa. With the iron yield per SNIa of 0.74 \Ms,
$3.4\times 10^8$ SNeIa are required to produce the iron in the plume. We
discuss a few possiblities for the origin of the iron-rich plume.

Frequent minor-merger events are expected in the cD galaxy. The plume
could be the gas stripped from an infalling metal-rich galaxy,
although iron needs to be selectively left behind. More likely is star
formation taking place in the gas deposited by a merger. A short
duration starburst results in a metallicity pattern of SNeIa after a
significant number of SNeIa start to explode (e.g., Hamann \& Ferland
1993; Yoshii, Tsujimoto \& Kawara 1998). However, an immediate
difficulty is the lack of excess in optical light there.  The $\sim
10^{10}$ \Ms\ or more of star formation for producing the required number
of SNeIa should be visible in blue light. There are two other potential
problems, depending on evolutionary models of SNIa
progenitors. Firstly, even if the progenitors were formed in a clump,
they would have moved in the galactic potential and phase mixed within
a timescale of order a few times the crossing time of the plume $\sim
5\times 10^7$ yr (its size, 14 kpc divided by the orbital velocity of
say 300 \kmps). This is consistent only with the lifetime of currently
disfavoured doubly-degenerate (DD) progenitors (Tutukov \& Yungelson
1994) and single degenerate (SD) with a large mass (up to 8\Ms)
secondary star (Matteucci 1994), and not with that of SD binaries with
a smaller mass secondary star (SNeIa only start to explode after 0.5
Gyr for an elliptical, Kobayashi et al 2000). Secondly, the time scale
of SNeIa enrichment must match the short cooling time of the gas. For
an elliptical galaxy with $L_{\rm B}\sim 10^{11}$\Ls\ like 4C+55.16,
the peak SNIa rate, which lasts for less than 1 Gyr, is expected to be
0.1--1.5 SNeIa yr$^{-1}$ (Kobayashi et al 2000; Matteucci 1994). With
this range of SN rate, it takes 0.2--3 Gyr to produce $3.4\times 10^8$
SNeIa, hence the iron mass in the plume. Thus the cooling time of the
plume requires a SN rate at the higher end of the expected range to be
matched by the iron enrichment time scale. It should also be noted
that a similar amount of iron is contained outside the plume within
the central 25 kpc.

An alternative possibility is that the plume has been ejected from the
centre of the galaxy. Gas enriched by SNeIa may be retained within the
steep cluster potential until ejection takes place. However, in
order for the gas not to cool and then drop out, some form of heating
to combat the radiative cooling is required. SN heating can only
partially offset the cooling; the thermal energy in the plume is
$(3/2)n_{\rm e}kTV\approx 10^{59}$ erg, which is equivalent to the
total energy from the required $3.4\times 10^8$ SNeIa, assuming
$10^{51}$ erg per SN and no SN energy is radiated. Using the cooling
time as a guide, the projected distance of 14 kpc between the plume
peak and the centre of the galaxy requires the ejected gas clump to
travel at a velocity of $\geq 100$ \kmps. Since no sign of a temperature
rise in the ICM due to a shock is seen, the motion of the galaxy must be
subsonic, i.e., the velocity should be less than the sound speed of
$\sim 600$ \kmps\ for the 3 keV intracluster gas. At a velocity in a
range of 100--600 \kmps\ and the ICM density of $\sim 0.1$ \pcubcm,
whether ram pressure stripping works effectively is not trivial.
Although a dynamical study of the cluster is required to be
conclusive, from an inspection of the X-ray and radio images, the cD
galaxy appears to be lying at the centre of the cluster potential and
the radio lobe which perpendicular to the plume does not show any sign
of the radio-emitting plasma being swept back, indicating no evidence
for fast motion of the cD galaxy (at least on the plane of the sky).
One last possiblity is that the metal-rich plume is a fossil from when
the cluster core was hotter. This requires that mixing of metals with
surrounding gas is suppressed, much as conduction is reduced
(e.g., Vikhlinin et al 2001; Ettori \& Fabian 2000)

In summary, the high iron metallicity found in the
plume region is unusual when compared with well-studied nearby
clusters. Although the amount of iron could be supplied by SNeIa within
the central galaxy, how it accumulated is not clear. 
%Such
%an inhomogeneous metallicity distribution in the intracluster medium, if
%present on small scales in many clusters, could cause anomalies in the
%spectra of cooling flows, as discussed by Fabian et al (2001).

\section*{Acknowledgements}

We thank all the members of the Chandra team. 
The adaptive binning software was provided by Jeremy Sanders. 
%Ken'ichi Nomoto and Takuji
%Tsujimoto are thanked for useful discussion. 
The Royal Society
(ACF,SE,SWA) and PPARC (KI) are thanked for support.


\begin{thebibliography}{}

\bibitem{} Anders E., Grevesse N., 1989, Geochim. Cosmochim. Acta, 53, 197
\bibitem{} B\"ohringer H., et al, 2001, A\&A, 365, L181
\bibitem{} De Grandi S., Molendi S., 2001, ApJ, 551, 153
\bibitem{} Dickey J.M., Lockman F.J., 1990, ARAA, 28, 215
\bibitem{} Dupke R.A., Arnaud K.A., 2001, ApJ, 548, 141
\bibitem{} Ettori S., Fabian A.C., 2000, MNRAS, 317, L57
\bibitem{} Fabian A.C., et al, 2000, MNRAS, 318, L65
\bibitem{} Finoguenov A., David L.P., Ponman T.J., 2000, MNRAS, 544, 188
\bibitem{} Fukazawa Y., Makishima K., Tamura T., Ezawa H., Xu H., Ikebe Y., Kikuchi K., Ohashi T., 1998, PASJ, 50, 187
\bibitem{} Fukazawa Y., Makishima K., Tamura T., Nakazawa K., Ezawa H., Ikebe Y., Kikuchi K., Ohashi T., 2000, MNRAS, 313, 21
\bibitem{} Hamann F., Ferland G., 1993, ApJ, 418, 11 
\bibitem{} Hutchings J.B., Johnson I., Pyke R., 1988, ApJS, 66, 361
\bibitem{} Ikebe Y., Makishima K., Fukazawa Y., Tamura T., Xu H., Ohashi T., 1999, ApJ, 525, 58
\bibitem{} Iwasawa K., Allen S.W., Fabian A.C., Edge A.C., Ettori S., 1999, MNRAS, 306, 467
\bibitem{} Kaastra J.S.,Ferrigno C., Tamura T., Paerels F.B.S., Peterson J.R., Mittaz J.P.D., 2001, A\&A., 365. L99
\bibitem{} Kobayashi C., Tsujimoto T., Nomoto K., 2000, ApJ, 539, 26
\bibitem{} Lawrence C.R., Zucker J.R., Readhead A.C.S., Unwin S.C., Pearson T.J., Xu W., 1996, ApJS, 107, 541
\bibitem{} Matteucci F., 1994, A\&A, 288, 57
\bibitem{} McNamara B., R., et al, 2000, ApJ, 534, L135
\bibitem{} Mushotzky R., Loewenstein M., Arnaud K.A., Tamura T., Fukazawa Y., Matsushita K. Kikuchi K., Hatsukade I., 1996, ApJ, 466, 686
\bibitem{} Nomoto K., Thielmann F.-K., Yokoi K., 1984, ApJ, 286, 644
\bibitem{} Nomoto K., Hashimoto M., Tsujimoto T., Thielemann F.-K., Kishimoto K., Kubo Y., 1997a, Nucl. Phys. A, 616, 79
\bibitem{} Nomoto K., Iwamoto K., Nagasato N., Thielemann F.-K., Brachwitz F., Tsujimoto T., Kubo Y., Kishimoto N., 1997b, Nucl. Phys. A, 621. 467
\bibitem{} Pearson T.J., Readhead A.C.S., 1981, ApJ, 248, 61
\bibitem{} Pearson T.J., Readhead A.C.S., 1988, ApJ, 328, 114
\bibitem{} Renzini A., Luca C., D'Ercole A., Pellegrini S., 1993, ApJ, 419, 52
\bibitem{} Sanders J.S., Fabian A.C., 2001, MNRAS, submitted
\bibitem{} Tutukov A.V., Yungelson L.R., 1994, MNRAS, 268, 871
\bibitem{} Vikhlinin A., Markevitch M., Murray S.S., 2001, ApJ, 551, 160
\bibitem{} Weisskopf M.C., O'Dell S.L., van Speybroeck L., 1996, Proc. SPIE, 2805, 2
\bibitem{} Whyborn N.D., Browne I.W.A., Wilkinson P.N., Porcas R.W., Spinrad H., 1985, MNRAS, 214, 55
\bibitem{} Yoshii Y., Tsujimoto T., Kawara K., 1998, ApJ, 507, L113
\end{thebibliography}
\end{document}